\documentclass[AMS,stixtwocol]{WileyNJD-v1}

\articletype{Article Type}%
\usepackage{picins}
\usepackage{graphicx}
\usepackage{ulem}

\usepackage{tabularx}

\newcolumntype{Y}{>{\centering\arraybackslash}X}

\received{21 April 2020}
\revised{xx month 2020}
\accepted{xx month 2020}

\journalname{ETRI}
\journallogo{Wiley}{\textbf{ETRI} Journal}
\jurl{wileyonlinelibrary.com/journal/etrij}
\graphicspath{{Author_Photo/}}

\begin{document}



\title{6G in the Sky: On-Demand Intelligence at the Edge of 3D Networks}\protect
\author[1]{Emilio Calvanese Strinati}
\author[2]{Sergio Barbarossa}
\author[3]{Taesang Choi}
\author[5]{Antonio Pietrabissa}
\author[5]{Alessandro Giuseppi}
\author[5]{Emanuele De Santis}
\author[4]{Josep Vidal}
\author[6]{Zdenek Becvar}
\author[7]{Thomas Haustein}
\author[1]{Nicolas Cassiau}
\author[2]{Francesca Costanzo}
\author[3]{Junhyeong Kim}
\author[3]{Ilgyu Kim}
\authormark{CALVANESE STRINATI \textsc{et al}}

\address[1]{\orgname{CEA-Leti, MINATEC Campus}, \country{Grenoble, France}}
\address[2]{ \orgname{Sapienza University of Rome, DIET}, \orgaddress{\state via Eudossiana 18, 00184 Rome, \country{Italy}}}
\address[3]{\orgdiv{Telecommunications \& Media Research Laboratory}, \orgname{Electronics and Telecommunications Research Institute}, \orgaddress{\state{Daejeon}, \country{Republic of Korea}}}
\address[4]{\orgdiv{Dept Signal Theory and Communications}, \orgname{Universitat Politecnica de Catalunya}, \orgaddress{\state{Jordi Girona 31, 08034 Barcelona}, \country{Spain}}}
\address[5]{\orgname{University of Rome ``La Sapienza'' and Space Research Group
 of CRAT}, \orgaddress{\state{Via Ariosto 25, 00185, Roma, Italy}}}
\address[6]{\orgdiv{Faculty of Electrical Engineering}, \orgname{Czech Tecnical University in Prague}, \orgaddress{\state{Technicka 2, 16627 Prague}, \country{Czech Republic}}}
\address[7]{\orgdiv{Wireless Communications and Networks}, \orgname{Fraunhofer HHI}, \orgaddress{\state{Einsteinufer 37, 10587 Berlin}, \country{Germany}}}

\corres{*Corresponding author: Emilio Calvanese Strinati, \email{emilio.calvanese-strinati@cea.fr}}


\fundingInfo{European Union in
the Horizon 2020 EU-Korea project 5G-ALLSTAR, GA no. 815323, by the Institute for Information \& communications Technology Promotion (IITP) grant funded by the Korea government (MSIT No. 2018-0-00175), and by Grant No. P102-18-27023S funded by Czech Science Foundation.}
\abstract[Abstract]{6G will exploit satellite, aerial and terrestrial platforms jointly to improve
radio access capability and to unlock the support of on-demand edge cloud services in
the three dimensional space (3D) by incorporating Mobile Edge Computing (MEC) functionalities on aerial platforms and low orbit satellites. This will extend the MEC support to devices and network elements in the sky and will forge a space borne MEC enabling intelligent personalized and
distributed on demand services. 3D end users will experience the impression of being surrounded by a distributed computer fulfilling their requests in apparently zero latency. In this paper, we consider an architecture providing communication, computation, and caching (C3) services on demand, anytime and everywhere in 3D space, building on the integration of conventional ground (terrestrial) base stations and flying (non-terrestrial) nodes. Given the complexity of the overall network, the C3 resources and
the management of the aerial devices need to be jointly orchestrated via AI-based algorithms, exploiting virtualized networks functions dynamically deployed in a distributed manner across terrestrial and non-terrestrial nodes.
}


%

\openaccess{This is an Open Access article distributed under the term of Korea Open Government License (KOGL) Type 4: Source Indication $+$ Commercial Use Prohibition $+$ Change Prohibition.}

\maketitle


\section{Introduction}\label{Intro}
Coverage is a critical key performance indicator (KPI) when deploying wireless networks. Up to 4G networks, most of the 
efforts  have  been focused on increasing link capacity while ensuring a sufficient  coverage in the two-dimensional (2D) plane. 5G with its multi-dimensional requirements adds more stringent constraints for, e.g., mission critical services with requirements on a low latency and  a high reliability (URLLC), massive amount of devices (eMMB), range extensions, and on Operational Costs (OPEX) of the communication infrastructure. 5G allows exploiting new opportunities by sharing the underlying infrastructure among isolated and self-contained networks through the concept of network slicing. Moreover, starting from the 4G-LTE all-IP architecture, the network offers communication coverage and integration of cloud support. Nevertheless, services were offered on a 2D-plane, and cloud services were conceived for data fetching/storage (over significant distances between data centers and the users connected) and to provide services (e.g., social media or instant messaging) to mobile internet users. Newly emerging 5G services ask for solutions going beyond this framework, including ubiquitous coverage/capacity availability and service scalability adapted to new use cases, application scenarios and traffic conditions\color{black}, which would be a tough challenge for the one-network-fits-all 4G-LTE architecture.\color{black}

While the availability of a good terrestrial coverage has become common in densely populated areas and regions, the underlying business model based on a flat fee per user does not scale well in sparsely populated regions or areas with difficult orography (e.g., islands, rugged mountainous terrain or off-shore). Worldwide mobile network operators provide usually no, poor or at best a low-quality connectivity in those cases, while the potentials of these regions can only be fully exploited when providing connectivity for the digitisation of their economic activities, e.g., smart agriculture or mining. Relevant KPIs in this context are ubiquitous connectivity, scalability, and affordability. Moving from 2D to 3D-coverage is an enabling solution, the third dimension resulting from placing network elements up into the sky and space.
\vspace*{-20pt}\subsection{\color{black} Cooperation among Terrestrial and aerial/spatial networks}
Many recent research projects investigate the cooperation between terrestrial and low earth orbit (LEO) satellite networks for 5G new radio (NR). Within the 3GPP framework,  use cases and associated system requirements for the satellite integration in the 5G eco-system are specified and continuously updated by the working group SA1 in \cite{3GPPTS22.261}.
Standardization impact to the NR specification are studied in \cite{3GPPTR22.822}\cite{3GPPTR38.811}, considering non-terrestrial networks (NTN) as an integral part of NR. The successful outcome from these studies led to normative work in Release 16 specifying extensions to NR for unmanned aerial vehicles (UAVs) \cite{3GPPTS22.125}, high altitude platform stations (HAPS) and satellites based on well defined channel models, deployment scenarios, and system parameters. Likewise, future 3GPP releases will focus on solutions for RAN protocols and architecture. 

The 5G AgiLe and fLexible integration of SaTellite And cellulaR (5G-ALLSTAR) H2020 project \cite{5GAllstarWCNC2020} investigates multi-connectivity technologies that integrate  cellular and satellite networks to provide reliable, ubiquitous and broadband services for 5G \color{black}NR\color{black}. This is the follow-up of first investigations on terrestrial with non-terrestrial communication integration in the 5G CHAMPION project \cite{Calvanese2018}. Multi-connectivity requires significant innovations in the integration of millimeter-wave (mmWave) 5G-NR-based cellular system with a NR-based satellite system, as well as adoption of spectrum sharing and interference management techniques. The H2020 project VITAL addresses the terrestrial and satellite networks by enabling Software-Defined-Networking (SDN) based, federated resources management in hybrid satellite-terrestrial networks. The H2020 project SANSA aims at enhancing capacity and resilience of wireless backhauling through the cooperation of terrestrial-satellite networks. In these projects load balancing, efficient spectrum usage, improved coverage and link performance are sought.

HAPS \cite{Arum2020} are unmanned aircrafts positioned above 20 km altitude, in the stratosphere, for very-long-duration flights counted in years. Since the 1990s, a number of initiatives have been launched worldwide to explore the potential applications, including telecommunications services. HAPS offer wide area coverage with advantages compared to satellites in terms of cost, ease of deployment/reuse and large payloads, lower delays and signal attenuation. Recently, the Google's Loon project has been deploying a network of high-altitude solar-powered balloons that move using wind jets. They embark regenerative payloads and inter-balloon communication links and its network coexist with terrestrial LTE networks providing service to rural mobile broadband users in areas where terrestrial coverage does not exist. Some other operational HAPS with higher payload capacity (like Thales-Alenia's Stratobus dirigible) are expected by 2021-2023.

 At a lower altitude, drones are UAV that have the capacity of dynamically providing radio on demand coverage exploiting embarked light base stations \cite{Chandrasekharan2016}\cite{Fotouhi2019}. UAV's and HAPS have received considerable attention \cite{Zeng2016} in terms of data traffic management \cite{Lyu2018}, network coverage enhancement \cite{Li2019a} \color{black}\cite{7962642}\color{black}, improving quality of service \cite{Wang2019} \cite{Plachy2019}, propulsion and transmission powers \cite{Zeng2019}, \color{black} latency minimization \cite{8533634}\color{black}, or exploitation of network access \cite{Nasir2019}. 

\vspace*{-20pt}\subsection{\color{black} Hierarchical BS fleets for providing computing and intelligence functionalities} 
\color{black} Several works in the literautre such as \cite{8533634, Zeng2019,Gapeyenko2018} and \cite{7962642} propose different architectures and mathematical models for 3D networks comprising multiple UAVs, focusing in particular on the communication aspects such as  data backhauling and reduced latency, whereas the architecture that will be presented in this paper focuses on joint communication, computation, and caching capabilities,  which are considered as components of a single 3D system. \color{black}
Extending the use of UAV's to provide not only radio access, but also mobile computing functionalities is actually considered a promising paradigm to satisfy {\it on demand} communication and computation requests, and deliver {\it context-aware} cloud services to mobile users.
The first attempt to host cloudlet processors on the UAV, is addressed in \cite{Jeong18}. The target is to minimize the energy at the UEs, while optimizing transmission data rates, jointly with the UAV's trajectory under latency constraints. In \cite{Xiong19} the authors include an edge computing scenario with aerial platforms and heterogeneous IoT devices. A dynamic formulation appears in \cite{Zhang19}, where computation offloading is handled with Stochastic Optimization tools having energy consumption as a goal while optimizing the trajectory of UAVs. In \cite{Costanzo20}, a dynamic online strategy jointly allocates communication and computation resources, while selecting the vehicle's altitude, with the aim of minimizing the system energy and satisfying latency constraints. The work in \cite{HOU19} introduces Fog Computing into swarm of drones, with the aim of handling computation-intensive offloading of tasks. In \cite{Yang19} the sum power consumption is minimized for a multi UAV-enabled MEC network.

In our vision, research is needed to investigate solutions in realistic scenarios in which 3D services are supported by a hierarchical fleet BS embarked in UAVs, HAPS and LEO satellites, each having its own specific features in terms of payload, flight autonomy, mobility, service coverage time, altitude, revisit time, computation, storage, coverage area, link power budget, etc. In such a challenging context, ensuring end-to-end service continuity for ground users or to users moving in the 3D space entails rethinking the mobility management mechanisms incorporating proactive allocation of the content, smart proactive caching of recurrent computational results \cite{DiPietro2019}, instantiation of virtual machines, interference management and joint handover between radio access points and mobile edge computing hosts. This will require the development of a fast live migration of the light virtual machines, e.g., dockers, and an extension of NFV/SDN orchestration schemes to make them more inclusive with respect to the types of network nodes and also faster to support the mobility of both user terminals and network elements. 

Artificial intelligence (AI) can help to solve these issues. The last decade has witnessed a rapid progress in the field, driven by the increased computational capacity of computers and the wide availability of data sets. In end-to-end communications, ETSI Experiential Network Intelligence (ENI) group investigate how 5G networks can leverage AI to achieve autonomous, and thus cost-effective, slice management and orchestration. Inspired by the success of AI in solving complicated control and decision-making problems, distributed AI approaches are enablers to allow the network functionalities to learn about the network and take the best decisions accordingly.


Looking into the predictions of new technologies and services for the next decade, there is a clear need to move beyond 2D service coverage to truly 3D native services. 6G networks will enable end users moving in the 3D space to perceive a surrounding “huge artificial brain” offering virtually zero-latency services, unlimited storage, and immense cognition capabilities \cite{Calvanese2019}. 
To match this vision, future 6G networks will seamlessly incorporate terrestrial, aerial, and satellite radio access points to teleport \textit{on demand} cloud functionalities \textit{where} and \textit{when} the intelligence support is needed in the 3D space.

\vspace{-15pt}{\subsection{Purpose and structure of the Paper}
The purpose of this article is to identify a set of  technological advances, to highlight the main research challenges and open issues for the next decade of research to move beyond pure 2D service coverage to truly three dimensional (3D) native service. In this direction, Section \ref{SystemModel} details the foreseen 3D hierarchical system architecture for future 6G networks. 
Section \ref{B5G-3D42D} presents the enhancement of 2D terrestrial connectivity and services towards 3D support of joint communication, computation, and caching (C3). }
%
%
Section \ref{3GPP_STANDARDIZATION} details the current status of standardization bodies, future trends in the integration of NTN for 5G \color{black}NR \color{black} and opportunities of innovation for 6G networks.
%
%
Section \ref{section:interference} evaluates solutions for interference management in 3D hybrid intelligent networks.
Section \ref{DynamicCAC} describes possible solutions for effective management of multi-RAT resources in the 3D space through dynamic admission control mechanisms and load balancing.
Key observations and concluding remarks are in section \ref{Conclusions}.


\section{System Architecture}\label{SystemModel}
\begin{figure}[t]
    \centering
    \includegraphics[width=\columnwidth]{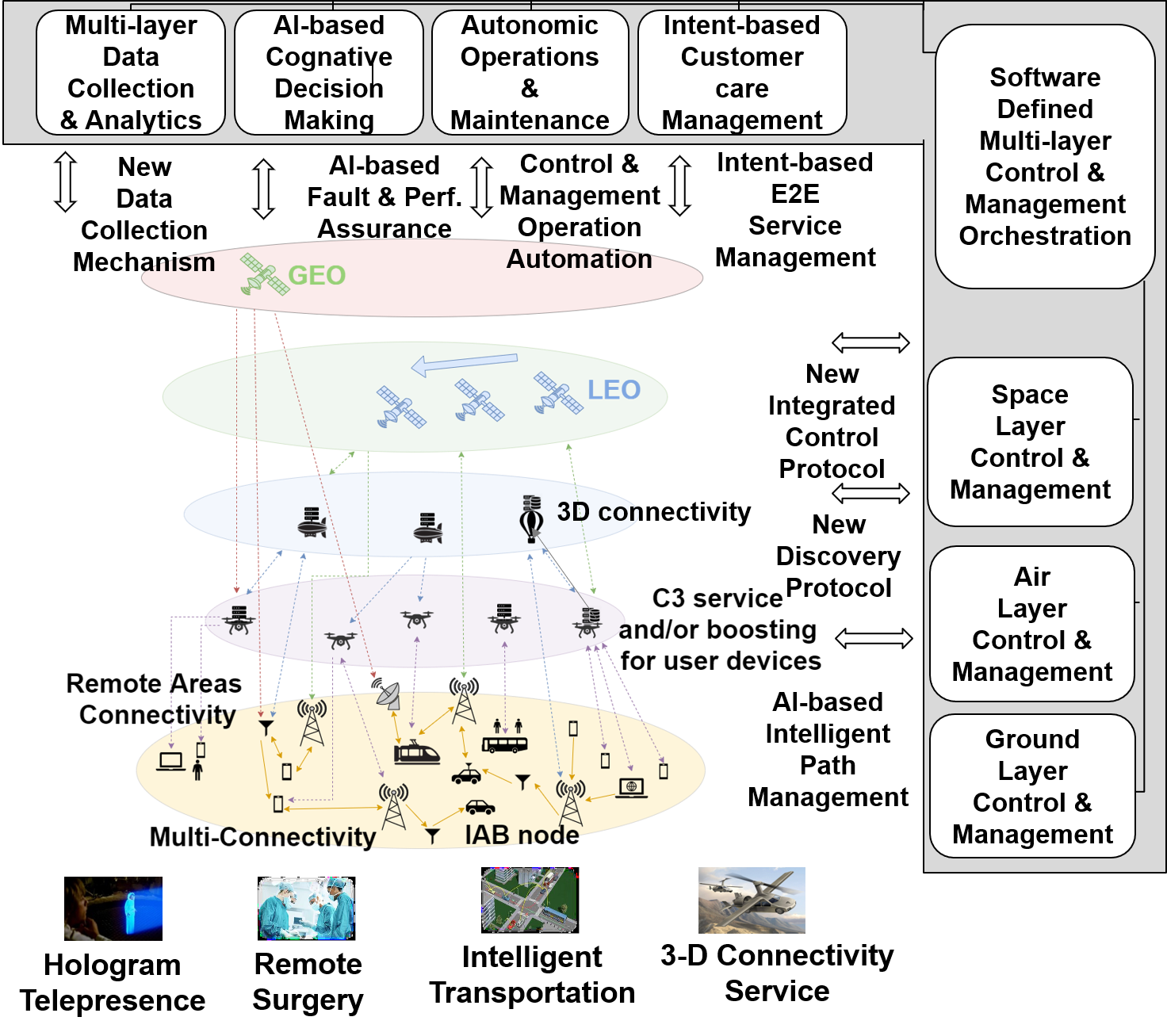}
    \caption{\color{black} Hierarchical 3D Network System Architecture}
    \label{fig:SysArch}
\end{figure}
Hierarchical 3D networks with multiple and heterogeneous types of flying layers are key to provide enhanced 2D services \cite{Ahmadi2017} and to 3D native services including connectivity and intelligence support.
Figure~\ref{fig:SysArch} illustrates a high-level architecture of the hierarchical 3D networks
unifying diverse 3D network nodes distributed over ground and flying layers.
Different types of aerial nodes such as UAV, and more generally Low Altitude Platforms (LAPs), HAPS and LEO/GEO satellites are located on different flying layers. Since aerial nodes can be equipped with on-board computation/storage capabilities, they can serve as 3D Base Stations, alone or in swarm formation, or 3D relay, which comprises an integrated access and backhaul (IAB) based hierarchical 3D networks. Although the current IAB standardization in 3GPP focuses on the ground network, in 6G, it will be extended to air and space networks as well as their integrated network. 

Low and high altitude platforms have several key potential applications in wireless communication systems due to their high mobility, flexibility, adaptive coverage capacity and low cost. Equipped with MEC servers, these aerial vehicles can provide opportunities for ground mobile users to offload heavy computation tasks, and then after computation, the mobile users can download the computation results via reliable, cost-effective wireless communication links, as well as download each kind of needed content. The proposed integrated 3D architecture enables the boosting of C3 performance in areas with existing infrastructure, and provide a network infrastructure for C3 services in areas without coverage. 3D connectivity services exploit the flexibility to accommodate a wide spectrum of applications ranging from two-way telecommunications (e.g., interactive 3D video, 3D intelligent services), to remote sensing,  pollution monitoring, meteorological measurements, real-time earth monitoring, traffic monitoring and control, land management and agriculture.

Connectivity of UEs, BS and relays placed on different flying layers might lead to much larger connectivity handover instances, mainly due to the difference in heights and speeds of nodes belonging to different flying layers. A today open axe of research for offering 3D service continuity and handover instance minimization are the cross-layer harmonization of selected UAV, HAPS and satellite placement and, the optimization of flying trajectories.
In addition, already in 2005 the NASA proposed the vision of a \textit{Space Wide Web network}, where messages can hops between intermediate nodes to reach close planets having each each orbiter, rover, space-borne telescope, and any other skyward-launched device working as a node of the 3D network \cite{Jackson2005}. At the horizon of 2030, with 6G, 3GPP standards will not go so far. Nevertheless, an \textit{Sky Wide Web or Internet of Sky} might be already possibly interconnected with 6G non-terrestrial 3D networks. 
In this hierarchical 3D network, 3D multi-connectivity will allow UEs to establish multiple different traffic links with 3D network nodes, thereby significantly improving service performance of the UE with a dynamic load balancing scheme over the established links. This, however, requires specifically designed highly efficient and intelligent control and management of 3D layers.

In our view, future 3D system architectures will apply network slicing  not only across terrestrial nodes as it is designed for 5G networks, but also across non-terrestrial nodes to facilitate different use cases and services provisioned in 3D space. \color{black} The proposed architecture shall then be able to offer services that go beyond pure connectivity and at the same time offer deep customization of connectivity and intelligent mobile network services at different granularity levels, spacing from dedicated slices per data of users, to slices per individual and groups of users and to slices dedicated to 3D applications and 3D sub-networks. This will require a new adaptable \textit{Midhaul} for an era of services that goes well beyond services of today 5G networks and the ones envisaged in most studies that focus on integrating UAVs into 5G networks. \color{black}


\color{black} AI-based approaches for network control also play a pivotal role in intelligent routing selection across 3D network layers and load balancing. For this reason, the proposed architecture shall be able to provide network intelligence capabilities at various levels and also  entail  device-to-device (D2D) communication, which may be enhanced by the addition of the new dimension and moving network equipment such as UAVs. \color{black}
In 3GPP, the first version of NR sidelink for the support of advanced V2X applications has been developed in Rel-16, 
and in 3GPP Rel-17, sidelink-based relaying functionality will be studied on top of the Rel-16 sidelink specification
for the purpose of sidelink/network coverage extension and power efficiency improvement.
In 6G, device-to-device (D2D) communications will be further extended to 3D layers, which could have great potentials in facilitating a wider range of applications and services such as the next-generation intelligent transportation services.

\vspace*{-10pt}{\section{Standardization Perspective}
\label{3GPP_STANDARDIZATION}
Terrestrial mobile telecommunication standards are grouped into generations  (1G, 2G, 3G, 4G, 5G) and the related 3GPP Technical Specifications (TS) documents with particular feature sets have evolved  with associated  release numbering e.g. Rel.8 being considered the first release of 4G-LTE and Rel.15 the first  of 5G-NR. 
Typically, a new generation arises at the confluence of significant maturity of new groundbreaking technologies and societal need for the introduction of new services that cannot be efficiently offered by current technologies. Standards define the set of new technologies to be included in the new generation. To this end, in order to have particular features becoming part of a standard release and/or specific technologies supported by the standard many aspects have to be considered. Beyond economic potential and technological maturity, it is very important to evaluate the standardization impact w.r.t. to the required changes to  existing specifications and features, which might already be rolled out in the market for billions of devices and base stations. The smaller the impact w.r.t. the Technical Specification and the bigger the expected commercial benefit  in the eco-system the higher are the chances for adoption into a standard.}

\vspace{-20pt}{\subsection{Terrestrial  and Non-Terrestrial Networks}
Since many years, researchers have been advocating solutions for  a converged integration of terrestrial and satellite communication into handheld devices and mission control centers \cite{Chandrasekharan2016}, which ranged   from Over-The-Top  multi-RAT approaches \cite{BATS_D2.4} to fully unified air-interfaces \cite{Jungnickel2012}. Conducted field trials with e.g. adapted  4G-LTE system parameters \cite{DOMMEL2014} proved feasibility but only recent advances in   5G-NR standardization \cite{3GPPTR38.811} finally bring commercial impact into graspable reach.}

Continuous efforts were made by the satellite community to engage and contribute in the 3GPP process, which was focused on land mobile networks for decades. The inclusion of NTN use cases and deployment options into the 3GPP technology feature roadmap is a best practice example on how vertical industries can actively push boundaries and get vital technologies  included into an evolving standard . 
Initial skepticism by many critics were overcome by a gradual approach, first to study the impact from NTN use cases on 5G-NR and to provide suitable channel models \cite{3GPPTR38.811} and simulation assumptions \cite{3GPPTR38.821} matched to the well-established 3GPP evaluation procedures and, after successful completion, continuing with nominal work in Release 16 and 17.  

The 3D component is a new territory for network design, in particular when aspects and KPIs like e.g. coverage, capacity, reliability, interference and mobility are to be extended and evaluated in 3D. It is expected that providing ubiquitous connectivity in 3D will require significant changes in architecture, function placement and network node design beyond the current approaches for terrestrial 5G base stations and satellites deployed of launched today. One example is MEC placement in a LEO satellite network to provide, e.g., a virtual private network slice for maritime or air fleet applications with low latency service requirements.  MEC placement may require fundamentally new approaches of dynamic allocation of computation, caching and communication resources on LEO nodes, including inter-node connections in space and between space and ground. Thus, the standardization impact goes beyond 3GPP and will touch standardization groups in charge of MEC, SDN, Fronthaul, and other interfaced involved to build a fully functional communication network.  


Latest  satellite network deployments will increasingly populate the low earth orbits (LEO) at 500-1000 km altitude.  Various corporations and consortia e.g. Amazon's Project Kuiper, OneWeb, Telesat or Elon Musk's Starlink
plan to provide internet services from 2021, with current deployments ranging from a few dozens to hundreds of  satellites, some targeting more than 10.000 in the future.  Bend-pipe satellites keep flexibility for air-interface selection,  e.g. DVB S2X or LEO adapted variants of LTE, NB-IoT or 5G-NR.  On the other hand on board signal processing  help to reduce e2e latency, in space packet routing and MEC. This will further open the existing satellite ecosystem toward interoperability and scalability in market size on chip, module, device and signal processing platform manufacturers, system and service provider level.
Since satellite networks provide coverage footprints beyond boundaries of countries or continents, infrastructure and spectrum sharing  will become increasingly important for cost and spectrum efficient deployment and operation of terrestrial and NTN including VLEO, cube sats and HAPs.

\subsection{UAVs as machine type  devices}
For many years, Unmanned Areal Vehicles (UAV) were used for research and tactical applications and  remote controlled mini drones had to follow line of sight (LOS)  constraints within a few hundred meters. With sufficient coverage footprint, mobile networks can enable new UAV use cases treating drones like user equipment (UE). As a consequence, the coverage footprint has to be extended reliably into 3D where cross-link interference becomes an issue since LOS links between a UAV and many base stations were not  considered in  cellular network design so far.  Similar considerations apply for connecting airplanes from ground using 4G or 5G base stations \cite{Chandrasekharan2016}, where the  flight corridors for civil aviation are  sufficiently well separated from the close-to-ground drone traffic.
Release 15 3GPP has studied \cite{3GPPTR36.777}  the ability for UAV  to be served by LTE and NR networks, identifying further performance enhancements for UE- and Network-based solutions, DL and UL interference mitigation, mobility performance and aerial UE Identification. Further enhancements \cite{3GPPTS36.331} addressed the issue of aerial UE interference  helping the eNodeB to see the UAV and to deal with any potential interference. Release 16  specified mechanisms for  Remote Identification of UAVs \cite{3GPPTS22.125} and   3GPP SA1 completed a study into requirements and use cases for the services to be offered, based on remote UAS identification, which are to be continued  in Rel.17 \cite{3GPPTR23.754}.
With the given maturity of UAV support in the 3GPP standard, UAV operation and/or assistance over cellular networks is becoming close to commercial deployments aside from regulatory constraints which seem to be changed in many regions in exploratory steps.
Full scale deployments of UAVs for use cases like parcel delivery are  years ahead providing sufficient time from early field trials and commercial roll-outs to feed back into the standardization process for 5G+ and 6G.

\vspace*{-10pt}{\subsection{UAVs as Radio Access Network}
Instead of connecting UAV with an existing RAN for control and communication from on-board equipment and/or sensors, UAVs may serve as deployed base stations or provide relaying functionality between devices and base stations of the RAN.
Prominent examples of flying base stations for emergency networks or networks in remote areas  are Google's Loon project \cite{Google_Loon} or unmanned airplanes supporting a larger coverage area while moving above the targeted coverage area at an altitude of 10-20km. Alternative approaches consider drones at very low altitudes of a 10-50m to provide extra capacity at hotspots \cite{Fotouhi2018} e.g. during large public gatherings.
Considering non-stationary positions and a varying number of infrastructure components e.g. UAV mounted base stations to provide an extended cellular coverage, such dynamic topology with all its flexibility comes at the cost of additional features at the RAN side to be standardized.  So far, moving base stations and/or networks have been tested and deployed in relative isolation, using proprietary interfaces in particular for backhaul and interlinking between several base stations using LOS-links over potentially hundreds of kilometers with mmWave or laser technology. For a wider acceptance in co-existence  with  terrestrial RAN deployments further studies have to be made beyond the ongoing discussions for 5G-NR. }

\vspace*{-10pt}\subsection{MEC Mobility: ETSI}
The framework and reference architecture for multi-access edge computing (MEC) was specified in ETSI \cite{ETSI_GS_MEC} and provides interfaces and messaging for integration and orchestration within a RAN specified  by 3GPP for 5G-NR and IEEE for e.g. 802.11ac/ay and used for feasibility studies \cite{MiEdge2017}. 
The dynamics of NTN topologies and the fact that we have a network of moving nodes has impact on the existing standard w.r.t. predictive MEC Handover, user group handover and session migration as well as meshed backhaul and multi-connectivity for mobile access.


\vspace{-10px} \color{black}\section{From 5G NR 2D Enhanced Services to 6G 3D Services}\color{black}
\label{B5G-3D42D}
In this section, we focus on the coverage extension from 2D to 3D. First, we analyze the benefits of the inclusion of the aerial devices in terms of connectivity. Then, we move to the service level, highlighting the need for moving to a holistic approach that looks at communication, computation and caching as components of a single system. We distinguish between 2D services involving devices on the ground potentially benefiting from 3D connectivity, and 3D services involving devices on the ground and in the air. We discuss these aspects also from the point of mobility management, handover and live migration of virtual machines and, control of C3 services. Finally, we focus on the importance of including artificial intelligence mechanisms to design a cost-effective system, able to incorporate proactive mechanisms and to learn from online observations. 

\vspace{-10pt}{\subsection{3D Connectivity}
Including UAV-based devices in wireless communication networks provides a cost-effective solution to improve connectivity, especially if the data traffic is non-homogeneous and non-stationary, i.e. it is expected to be highly varying across space and/or time. In such a case, a fixed infrastructure is highly ineffective, for both CAPEX and OPEX expenditures. As in many real world situations, the opportunities offered by the UAV-based devices come along several challenges. To highlight these challenges, it is first necessary to classify the role that UAV-based devices can play in the network. The UAV devices may act as: flying bases stations (UAV-BS), as flying user equipments (UAV-UE), and as flying relays (UAV-R). }

The UAV-BS brings connectivity to the mobile devices on demand. The challenges come from the nature of the UAVs. HAPS have sufficient energy availability and are typically supported by solar-powered batteries, so that they are able to support continuous coverage for a long time. They can typically be used to support long term coverage purposes. On the contrary, the support of coverage in highly time-varying situations is better handled with LAPS, which can be flown on the spot of interest on demand. However LAPS have very limited energy availability and can hover over a given area for a relatively short period of time. This means that flight and energy constraints should be taken into account in allocating the resources of the network. The limited weight payload that can be placed on a LAP suggests the use of higher frequency bands, e.g., mmWave bands, to use smaller size antennas and to achieve better spectral efficiency. However, the use of mmWave links faces the problems of link attenuation, in case of rain, and blocking effects. To reduce link attenuation it is necessary to limit the coverage area, possibly flying at the lowest permitted altitude. However, flying at low altitudes increase the probability of blocking. Momentary blocking  severely impacts also the reliability of high capacity radio links and the MEC assisted service-continuity \cite{Barbarossa2017}. In 2D networks, the detrimental effects of blocking are reduced using multiple Radio Access Technologies (RATs) or multiple interfaces of the same RAT. The adoption of 3D connectivity to enhance performance of 2D networks brings interesting new opportunities and challenges to be solved \cite{Zeng2016} for the next decade in 5G and beyond 5G networks. The selection of the altitude plays a key role. Intuitively, the higher is the altitude, the larger is the coverage offered by the platform and the lower is the chance of suffer shadowing effects, due to favorable Line-of-Sight (LoS) propagation conditions. However, high altitudes also imply larger distances and then higher attenuation. 
The altitude has then to be carefully selected, also depending on the distribution of the UE's \cite{Costanzo20}.

To enable the several applications of UAV-assisted services, the UAV's need to communicate with the existing wireless network, either cellular or Wi-Fi. In such a scenario, the UAV's act as the UEs of the wireless networks. The UAVs can also act as UEs in applications such as delivery drones, real-time surveillance and UAV-assisted transportation networks. In this case, we have a really {\it 3D service exploiting a 3D network architecture}. An interesting example of 3D service is a virtual reality scenario, where the UAV flies over a location of interest carrying a 360 degree camera, which is controlled from the end user equipment to select the view angle specifying which part of the video needs to be transmitted with sufficient quality. To handle these 3D services properly, it is necessary to handle the interference that UAV-UEs can bring to the terrestrial UEs. Typically, the antennas of current terrestrial BSs are designed to handle an essentially 2D coverage problem, so that the radiation patterns is usually attenuated at high elevation angles. As a consequence, the communication between UAV-UEs and conventional BSs typically relies on sidelobes or back lobes of the BS antenna. Clearly, a better design involves a proper redesign of 3D beamforming at the BS, able to track the UAV-UEs. In \cite{3GPPTR36.777}, 3GPP specifies new BS antenna design and cellular communication techniques for UAV coverage up to the maximum altitude of 300 m. Most likely, it will be necessary for the BS to distinguish between the aerial and terrestrial UEs, to handle them separately. 

Finally, UAV devices can act as relays (UAV-R) to provide backhaul the connectivity between the terrestrial/aerial UEs and the terrestrial/aerial BSs. In such a case, key challenge is to devise effective cooperative communication strategies that take into account the mobility of the aerial devices. In principle, one could make near distance UAV devices operate as a huge virtual antenna, with also the possibility to adapt the shape of the constellation by making the UAVs to move as needed, provided that the resulting synchronization problems are properly handled. In general, using the UAVs as wireless relays can boost (on demand) the link quality between the ground BSs and the terrestrial UEs, but it raises also an interference issue towards the neighboring BSs that should be handled consequently. 


\vspace*{-10pt}\subsection{C3 Support Extension}
The UAVs can be used not only to improve connectivity, but also to bring (cloud) services closer to the end user on demand, thus, extending the concept of edge computing, or fog computing, to incorporate the aerial devices as the edge of the network. In this way, delay-sensitive services can be delivered where and when needed \cite{Calvanese2019}. Of course, the flight and energy constraints of some aerial devices, such as LAPS for example, need to be incorporated in the system design. 
As an example of application, in the IoT scenario multiple sensors send their data for processing and detection of possibly anomalous situations. Flying the UAV-BS close to these sensors, having sufficient computational capability, can be very effective to implement a computation offloading strategy able to extract relevant information from the data near the location, where the data is collected, thus accommodating for strict delay constraints. In this way, the UAV-BS can help sensors to run (remotely) sophisticated algorithms or to prevent an excessive energy consumption. In other applications, such as in disaster recovery, it is useful to bring content on demand in the areas where this context-aware information is needed. Also in this case, flying the UAV-BS with caching resources around the disaster area may be beneficial. \color{black} We also mention that 3D networks enhance the capabilities of Device-to-Device (D2D) communications, a technology in which UEs communicate with one another to exploit and share their resources that is studied in both emergency scenarios \cite{Cheng2019} and normal operations, such as decentralised and distributed computing  \cite{Ferrer2019,Mehrabi2019}. In fact, a moving fleet of UAVs may be also used for coordinating such D2D connections and enhance their coverage. \color{black} 

In other scenarios, when the computation requests cannot be handled even by the UAV, it can still be useful to fly the UAV close to the end user and to let it act as a relay to enable computation offloading to terrestrial devices that could be otherwise more difficult to reach, within the required delay constraints. In all these cases, since the overall delay incorporates the delay of (round-trip) data transmission from the UE to the UAV, the delay associated to computation, and possibly the (round-trip) communication delay between the UAV and the terrestrial edge cloud, it becomes clear that C3 resources should be handled {\it jointly}. Indeed, in a 3D network scenario, the C3 resources should also be managed together with the control of the UAV position, taking into account battery level on the UAV and the period of time in which the UAV can hover over the location of interest. This vision calls for a joint optimization of resources for control and C3 services. 

This vision calls for a very flexible orchestration of control and C3 resources, building on the virtualization of many functionalities. In this way virtual machines serving different purposes can be moved around to minimize the service delay and use resources only when needed and then release them when not. This requires the design of fast live migration of 'light' virtual machines across moving nodes of the same network tier or cross-networks tiers.

\vspace*{-10pt}{\subsection{Intelligent Handovers for Handover of Intelligence at the Mobile Edge of 3D Networks}
The handover of the UAV-BSs to the ground BSs becomes a part of the ground UE handover management problem. In such a case, both the ground UEs as well as the UAV-BSs compete for the same radio resources available at the ground BSs and, at the same time, the UEs can be also associated to the UAV-BSs. 

The mobility of all types of UEs is typically managed by individual BSs and the serving BS is responsible for monitoring and control of the UE’s handover. Nevertheless, with the introduction of computing capabilities (e.g., edge computing) in the network, the handover control functionalities can be deployed at any node having computation capabilities or even distributed across multiple nodes including UAVs in 3D space. The UAVs nodes are, however, often limited in terms of available energy. Thus, the control functionalities and the computation handled by these nodes in 3D can be handed over (or migrated) to another node over time to jointly optimize  communication, computation, and control functionalities’ deployment with respect to the energy currently available at these nodes. Thus, handover of communication and/or computation might lead also to handover of the control functionalities among the network nodes and vice-versa. This calls for a completely redesigned handover management incorporating not only communication, but also real-time control and computation aspects managed jointly with the possibility to associate the users to the UAV-BSs in order to jointly-manage seamless handover of communication, control, and computation in 3D.}

\vspace*{-18pt}{\subsection{Pervasive and On Demand Distributed AI at the 3D Edge}
The joint management of C4 resources \color{black}(i.e. C3 with the addition of control) \color{black} requires prior knowledge of many parameters of interest, like channel state, interference level, computation, and content requests, UAV navigation data, etc, to be able to run dynamic optimization algorithms necessary to allocate resources in order to satisfy the end-user requests, especially in delay-constrained services. However, typically most of these parameters are not known, or are only known in imperfect or outdated form. In this scenario, it is of paramount importance to resort to AI mechanisms to learn the unknown parameters from past observations and to predict the behavior of parameters of interest, to enable proactive resources allocation strategies, especially useful when dealing with delay-constrained services. A recent survey on the application of machine learning tools in a 3D scenario involving the UAV-based networks is \cite{bithas2019survey}. Machine learning algorithms, including supervised, unsupervised and reinforcement learning mechanisms have been developed to learn physical layer parameters, such as channel state and interference level, and to predict the received signal strength (RSS) at UAV's side. The user association problem, from multiple UEs to multiple UAV-BSs is formulated as a clustering problem, solved using simple k-nearest neighbor algorithms. Several works also address the challenges associated to the extension of the edge cloud to UAV-assisted devices. In \cite{hu2019ready}, a new method for the UAV clustering is proposed to enable efficient multi-modal multi-task offloading. Content caching on the UAVs is also proposed in \cite{chen2017caching}, where the authors exploit user-centric information related to content request distribution and mobility patterns for deploying the UAVs and for determining content caching on their buffers. Reinforcement learning mechanisms are well suited in the dynamic scenario modeling 3D communications. However, they may suffer from slow convergence because typically they start with no prior assumption. To speed up online learning, it may be beneficial to resort to stochastic optimization or online convex optimization mechanisms, extending the approach of \cite{chen2019learning} to the 3D communication scenario.}

\vspace*{-20pt}{\section{Interference Management for 3D Hybrid Intelligent Networks}\label{section:interference}
Hybrid 3D networks present a strong potential for enhancing service performance and reliability in the 2D plane and for enabling innovative services in the 3D space. Nevertheless, this is just a prospective gain. Interference can strongly limit such benefit and effective solutions for managing them have to be designed, implemented and validated. The common understanding is that interference, classically processed as additive noise, compromises the transmission and therefore must be ideally avoided or at least strongly limited.
Recent works model the impact of UAV and satellite generated interference on 2D communications. In \cite{Bergh2016} it is shown how interference is going to be a major limiting factor when terrestrial networks benefit from UAV support and how density of UAVs may generating rude inter-cell interference, causing catastrophic performance degradation. 

The 5G-ALLSTAR project investigates refined channel models and interference mitigation solutions for terrestrial communication enhanced by satellite links \cite{5GAllstarWCNC2020}. It will be crucial for a 3D multi-RAT systems 
to transmit with high out-of-band rejections to dynamically take advantage of any available spectral resource, with limited guard bands. We show in figure \ref{fig:results_adjacent} an evaluation of the performance degradation on a satellite link due to interference caused by terrestrial BS transmitting in 
adjacent channels. We evaluate the impact of the guard band and of the relative interferer power on the packet error rate (PER) for the K-band. We compared performance for three waveforms that can be demodulated with a 5G compliant receiver: Cyclic Prefix (CP)–OFDM and two filtered waveforms: Filtered (F)-OFDM and Block Filtered (BF)-OFDM \cite{demmer2017block}. 
It can be noted that the use of BF-OFDM makes it possible to avoid the insertion of large guard bands, even when the interference is \textit{strong}. In the case simulated here, the band gain is larger than 5$\%$.

\begin{figure}[t]
    \centerline{\includegraphics[width= 0.75 \columnwidth]{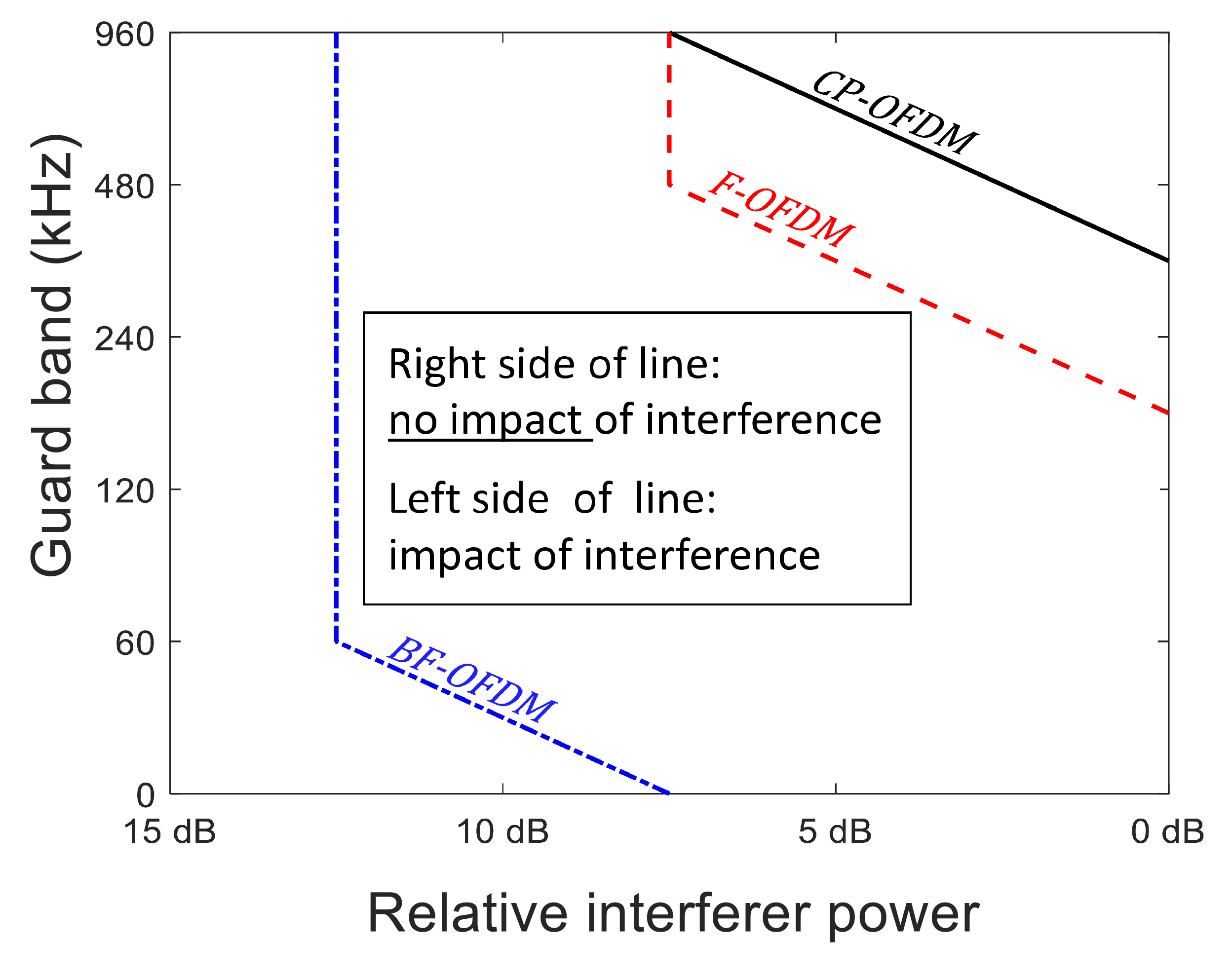}}
    \caption{Impact of terrestrial interference on satellite link (based on satellite PER). Satellite bandwidth: 15.36 MHz.}
    \label{fig:results_adjacent}
\end{figure}}

Results in \cite{5GAllstarWCNC2020} show also that non-terrestrial generated interference is either \textit{strong} compared to the intended signal or \textit{weak}. This result is also valid for 3D hybrid networks in which interference depends on  many factors such as UAV elevation and azimuth between receivers and interfering transmitters, side lobe gain, beam width, distance of transmitter and receive pairs, UAV volumetric density, etc. 
This opens novel opportunities for the design of innovative interference management techniques in which interference is not considered as an opponent but as a potential ally \cite{DeMari2014}. 
Recent works investigate how Coordinated Multi Point (CoMP) transmission and reception techniques applied to terrestrial communications can be optimized to limit the complexity of CoMP techniques. In \cite{Tran2017} it is shown how  a two-layer interference cancellation strategy assisted by MEC can notably limit useless CoMP associated processing for users that are in the \textit{weak} interference regime. Looking at the future, applying such concepts to clusters of non-terrestrial BS might require intense exchange of information to cancel the intra-cluster interference. Nevertheless, since inter-cluster interference is not addressed by CoMP techniques and 3D backhauling link might be unreliable, in case of dense clusters system capacity improvement can be negligible. Moreover, CoMP requires a large exchange of information and computational intense support of MEC. This might limit the NTN cluster size and indeed its potential support for 2D networks and pure 3D services. 

From our perspective, there is a strong potential for mitigating performance degradation caused by interference in and to NTN. We advocate that in such a context, ignoring or allocating resources for interference avoidance is not always the best option. In our view, in future 3D hybrid 6G networks, interference management techniques should exploit \textit{interference classification} techniques, following the concepts introduced in \cite{TSE} and then further evolved in \cite{DeMari2014} by the introduction of \textit{interference perception} concept and the simplification to a low-complexity \textit{2 regime interference classifier} which has only 2 admissible interference regimes for each user: the \textit{weak} and \textit{strong} regimes for which close to optimal solutions can be designed. While the benefits of IC techniques can notably improve performance, due to the associated complexity and to the many parameters to estimate, effective IC hardware implementations are still an open issue. 
We strongly believe for future 6G 3D networks in the potential of interference detection (possibly using ML techniques), combined with multi-layered CoMP \cite{Tran2017} with \textit{2 regime interference classifier} \cite{DeMari2014} and to consider trade-offs caused by practical implementation limitation of IC techniques,  coordination signalling and 3D MEC processing associated energy and latency costs. 
%
\vspace*{-20pt}{\section{Dynamic Resource Management for 3D Connectivity}\label{DynamicCAC}
\subsection{Multi-RAT CAC}
Mobile nodes acting as embarked relays or BSs in UAV can handle sporadic congestion events in the radio access network occurring in specific areas, by offloading communication and MEC traffic from the fixed terrestrial links (from the protocol stack viewpoint).

This scenario impacts on connection admission control (CAC) algorithms which now have to consider not only UE mobility, but also the mobility of the access point (AP). To show the capacities of this new scenario, a simulation study \color{black} made through an ad-hoc open-source 5G network simulator \cite{simulator_github} \color{black} is presented , in which the multi-RAT simulation environment is composed of one type of fixed RAT, provided by a satellite cell, and two types of mobile RATs: a 5G NR mobile relay node and a 5G NR mobile BS.

Besides the fact that the admission control must be capable of handling mobile APs in a 3D environment, the key of the success is the readiness of intervention. Based on traffic and mobility data, AI algorithms are needed to foresee when and where traffic peaks are going to occur for UAV to reach the identified area timely.
\vspace*{-20pt}{\subsection{Resource allocation}}
The resource allocation process differs depending on the RAT. For 5G NR RATs, we consider the Type 1 frame structure defined by 5G NR standards, which uses Frequency Division Duplexing (FDD) for both downlink and uplink, with minimum allocation unit defined as Resource Block (RB).  A RB is composed by 12 frequency subcarriers, whose bandwidth depends on the numerology $\mu$ \cite{3GPP2020-NR-mu}. The NR frame structure is composed by 10 ms frames, in turn composed by a number of time-slots depending again on the numerology $\mu$. Each RB is made by 12 or 14 OFDM symbols (respectively with extended and normal Cyclic Prefix).}


A different number of RBs is defined for each channel bandwidth, depending on the frequency band used (either FR1 \cite{3GPP2020-NR-FR1} or FR2 \cite{3GPP2020-NR-FR2}), and on the subcarrier bandwidth.






Once the UE requests a bitrate to an NR RAT, the AP computes the required number of RBs. Firstly, the AP computes the Signal-to-Interference-plus-Noise Ratio (SINR) for the UE. The inter-AP interference is estimated as

\begin{equation}
\begin{split}
	&\mathcal{I}_{ij} = \sum_{k\neq j}P_{i,k} \cdot \mathrm{RBUR}_k, \quad \textrm{with}\\
    &\mathrm{RBUR}_k = \frac{\sum_{\tau \in (t-T, t)}\sum_i I_{ij}N_{i,k}(\tau)}{T\cdot \mathrm{\#RB}}  
\end{split}
\end{equation}\\
being the Resource Block Utilization Ratio, where $I_{ij} = 1$ if UE $i$ is connected to AP $j$ and $I_{ij} = 0$ otherwise, $N_{ij}(t)$ is the number of RBs allocated to user $i$ by AP $j$ and $T$ is the time window for the computation of the average $\rm{RBUR}$. The data rate which can be transmitted by a single RB is computed using Best Modulation Coding Scheme (MCS) \cite{addali2019dynamic} with the Shannon formula: $r_{ij} = B_{\rm{RB}}\log_2(1+\rm{SINR}_{ij})$, where $B_{\rm{RB}}$ is the bandwidth of a single RB (i.e. $12\cdot 15\cdot 2^{\mu}$ KHz). Finally the number of RB needed to satisfy UE requirements is 

\begin{equation}
n_{ij} = \bigg\lceil\frac{R_i}{r_{ij}}\bigg\rceil 
\label{eq:n_rb}
\end{equation}

If the relative position between the UE and the AP changes, the SINR changes and the number of allocated RB to the UE has to be updated.

The simulated satellite RAT uses Time Division Multiple Access (TDMA) for concurrent UE access. Given a time frame, a certain number $n_{\rm{tot}}$ of symbols are available to the UE transmissions. Moreover, for each time frame, part of the symbols is used for synchronization purposes ($n_{\rm{sync}}$), each communication contains a header (of length $n_{\rm{head}}$) and there is a guard space of $n_{\rm{space}}$ symbols between each communication to avoid intra-RAT interference. The simulated satellite is an  Inmarsat implementation, with $n_{\rm{tot}} = 120832$ symbols, equivalent to a time frame of 2ms, $n_{\rm{sync}} = 288$ symbols, with 2 synchronization messages inside the time frame, $n_{\rm{head}} = 280$ symbols for each UE communication, $n_{\rm{space}} = 64$ symbols, $n_{\rm{slice}} = 39104$ symbols, that are about a third of the total symbols, $n_{\rm{block}} = 64$ symbols. \cite{maral2020satellite} The data rate that can be obtained by a single block is obtained from the Shannon formula, and the number of blocks to be allocated to satisfy the UE request $R_i$ is computed as in equation (\ref{eq:n_rb}). The actual integer number of symbols occupied by an UE are equal to $\bar{n}_{ij} = n_{\rm{head}} + n_{ij} + n_{\rm{space}}$.




\vspace*{10pt}{\subsection{Simulation Setup}
The environment is represented by a 4 Km $\times$ 4 Km grid containing 50 UEs, a single satellite AP and two mobile 5G NR APs. Each UE requires a bitrate of 10 Mbps, its starting position is randomly computed and it moves on a straight line with random direction at a speed of 10 m/s. We also consider the service is interrupted if the bitrate falls below 5 Mbps. The satellite AP is geostationary and uses a carrier frequency of 28.4GHz with 220 MHz bandwidth \cite{little2015high}. Its antenna Equivalent Isotropic Radiated Power (EIRP) is 62 dBW \cite{fenech2015high}. The path loss considers both the FSPL and the atmospheric loss (0.1 dB) and the user terminal G/T (-9.7 dB/K). The mobile 5G NR APs transmit a power of 15W, have an antenna gain of 15dB, a feeder loss of 1 dB and at 800 MHz carrier frequency with a bandwidth of 100 MHz and with numerology $\mu=2$.

The connection procedure consists in the following steps: (i) the UE measures the receiving power of the APs within its range; (ii) the UE chooses the AP to be connected to according to the received power with a User-Centric, RAN-Controlled or RAN-Assisted approach; (iii) upon communications with the UE, the AP allocates the resources based on the SINR in a best-effort basis. Due to the dynamicity of traffic and network elements, connection updates are required following the same procedure. The measured received power depends on the characteristics of the antenna of the generic AP j and on the path loss from the antenna to the UE $i$, i.e., $P_{ij} = P_j G_j L_j L_{ij}$, where $P_j$ and $G_j$ are the antenna power and gain, and $L_j$ and $L_{ij}$ represent the losses at the antenna side and the path loss between UE $i$ and AP $j$, respectively. The received power depends, via the path loss $L_{ij}$, on the relative positions of UE $i$ and AP $j$. The simulated path loss model of the satellite RAT is the free space path loss, whereas for terrestrial RATs (5G NR) we chose the COST-HATA \cite{cost-231}  path loss model. If a UE measures a receiving power lower than a threshold $P_{\rm{min}}$ for a certain AP, then the AP is considered not visible by the UE.

\setlength{\belowcaptionskip}{-10pt}
\begin{figure}
    \centering
    \includegraphics[width = \columnwidth]{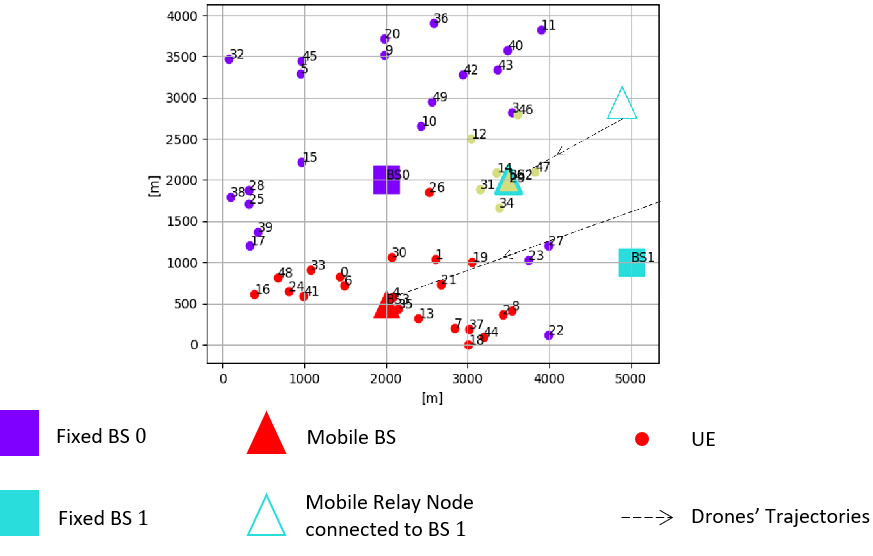}    
    \caption{Intervention of mobile BSs in the area covered by a BS and a satellite spot. The final position of APs and UEs is shown.}
    \label{fig:wp4_2b_map}
\end{figure}}

\begin{figure}
    \centering
    \includegraphics[width = 0.9 \columnwidth]{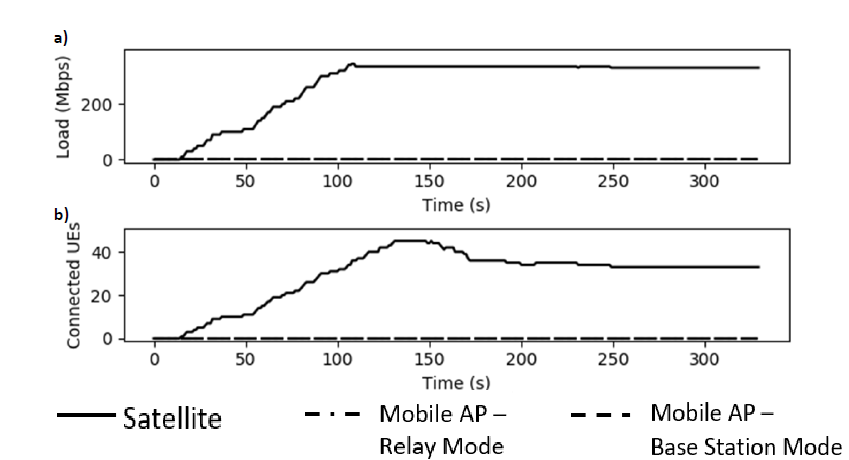}
    \caption{\color{black}a) Load of the satellite without the mobile APs. b) Number of UE connected to the satellite.\color{black}}
    \label{fig:wp4_2a}
\end{figure}

\begin{figure}
    \centering
    \includegraphics[width = 0.9 \columnwidth]{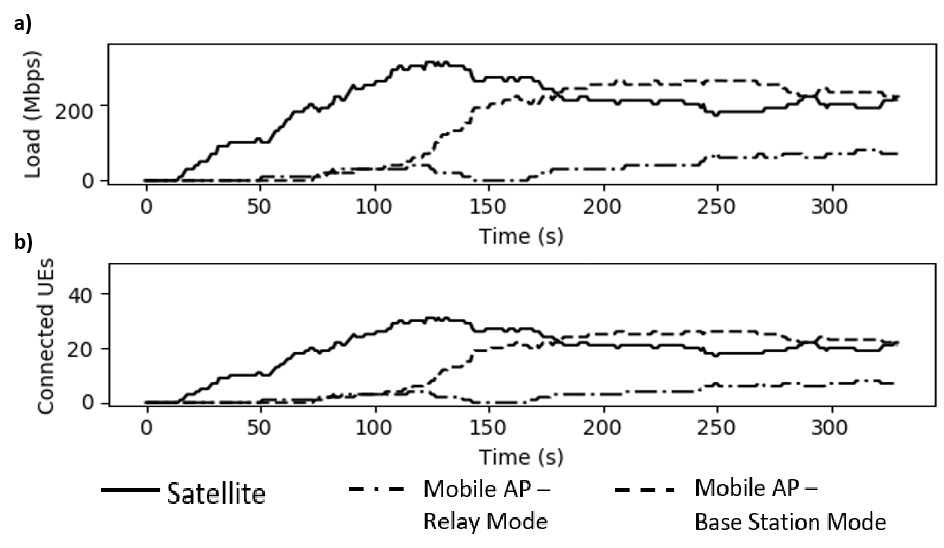}
    \caption{a) Load of the satellite and of the mobile APs. b) Number of UE connected to the satellite and to the mobile APs.}
    \label{fig:wp4_2b}
\end{figure}

\vspace*{-15pt}{\subsection{Simulation Results}
The simulation scenario shows how the use of mobile nodes can solve the congestion of fixed APs and assure \textit{service continuity}. The initial height of the two mobile 5G NR APs is 200 m and they are far from the UE positions, which in turn can only be connected to the satellite. 


The UEs start communicating at random time causing the satellite AP load to increase with time. With no mobile nodes available, the satellite AP eventually becomes congested, and new UE service requests are rejected, \color{black}as showed in Fig. \ref{fig:wp4_2a}\color{black}. Moreover, some of the UE bitrates fall below 5 Mbps, causing connections' drops and service interruption\color{black}, as in Fig. \ref{fig:wp4_2a}b\color{black}. On the contrary, if UAV APs are available, as in Fig. \ref{fig:wp4_2b}, the UEs start connecting to the mobile APs. In this case, no UE has to interrupt the service and service continuity is granted, maintaining the connections at 10 Mbps for the whole simulation time \color{black}as well as maintaining all the UEs connected to some AP\color{black}.}

\vspace*{-15pt}\subsection{Future research Directions for Dynamic Resource Management}
\color{black} We mention that this section reported only a preliminary simulation result, aimed at demonstrating the potentiality of a dynamic resource management algorithm for 3D connectivity and at the moment there are some limitations related to the 5G channel emulation, that focuses mostly on the estimation of the bitrate that could be sent over the radio link. Moreover there are limitations in the planning of mobile BSs trajectories in the 3D space to minimize the UEs connection interruptions. Various research activities are being carried out regarding dynamic network resource management proposing more advanced network controllers such as \cite{9143837}.
We also note that, the inclusion of non-terrestrial base stations brings new challenges in the network controller design \color{black}In fact\color{black}, as both the serving and interfering base stations can move at the same time in the 3D plane. \color{black} Another challenge is the prediction of user movements, that may avoid  excessive handovers. \color{black}

In parallel to the provisioning of connectivity to the ground users, \color{black}the connectivity of the non-terrestrial BSs to the core network should be guaranteed\color{black}. This connection can be provided at dedicated resources (different from user to ground base station) or at shared resources (\color{black}the same as used for common users\color{black}). The management of connectivity and handover of the non-terrestrial base stations among the ground base stations at the dedicated resources is, in its nature, the same as the conventional handover management of the ground users in the terrestrial-only mobile networks (such as 4\color{black}G\color{black}/5G). In such case, the non-terrestrial BSs compete for a connection to close-by terrestrial BSs only with others non-terrestrial \color{black}ones\color{black}. However, an allocation of the dedicated band for communication between the ground and non-terrestrial BSs is inefficient in terms of spectrum use. In contrast, the second approach with radio resources shared by all links from any BS to any user as well as to any non-terrestrial BS increases the spectrum reuse and enables more efficient exploitation of radio resources.

\vspace*{-20pt}{\section{Conclusions}\label{Conclusions}
The nature of new applications in the next decade and the desire of ubiquitous availability will most likely require technologies supporting truly 3D on demand services, rather than today's 2D service coverage.
In our view, while the integration of terrestrial with NTN for 2D service enhancement will come as a natural evolution of 5G, providing on demand connectivity and edge intelligence to support truly 3D services will not come before 6G.
In this article, we have provided an in-depth overview of a future hierarchical 3D network architecture where heterogeneous flying devices, providing different levels of mobility, coverage and service level, enable revolutionary new on demand connectivity and intelligence support.

Today, NTN use cases are already being considered for new features and technology extensions in the 3GPP standard releases 16 and 17. On the roadmap for 5G-NR, the integration of terrestrial and non-terrestrial networks will enable global 5G service enhancements and new functionalities. Beyond release 18 up to 6G, further extensions of 3GPP and other standardization bodies will enable advanced dynamic and meshed interconnection and relaying between NTN-nodes and MEC placement in the 3D space.

Some fundamental challenges remain open for future researches. We highlighted promising innovation directions, like on demand distributed C3 support, 3D-interference management, 3D-multi-link load-balancing and admission control, live intelligence handover and migration mechanisms, and AI-based joint orchestration of C4 distributed resources in the 3D-space.
%
%
Preliminary results, currently under investigation in the H2020 5G-ALLSTAR project, on interference management and 3D-multi-RAT admission control show that it is crucial for a 3D-multi-RAT system to transmit with high out-of-band rejection to dynamically take advantage of any available spectral resource. Recent results show that interference by non-terrestrial networks at the receiver is either perceived  as \textit{strong} or \textit{weak} compared to the intended signal. We advocate that this opens opportunities for the design of innovative interference management techniques in which interference is not considered as an opponent but as a potential ally.
Moreover, we show how additional 3D-nodes can effectively be exploited to dynamically handle network congestion, e.g., by using drones as on-demand mobile relay nodes or mobile BSs, to offload traffic from the fixed terrestrial links and/or to provide an extended opportunistic cellular coverage. In our opinion, new admission control procedures are needed to cope with the extended 3D network topology and, specifically, with the increased network handover occurrences implied by the dynamic 3D network.
}

\vspace*{-20pt}{\subsection*{Conflict of interest}
 The authors declare no potential conflict of interests.}


\vspace*{-22pt}{\bibliography{main.bib}}

\vspace*{-15pt}{\section*{Author Biography}}

\parpic{\includegraphics[width=60pt,height=70pt,clip,keepaspectratio]{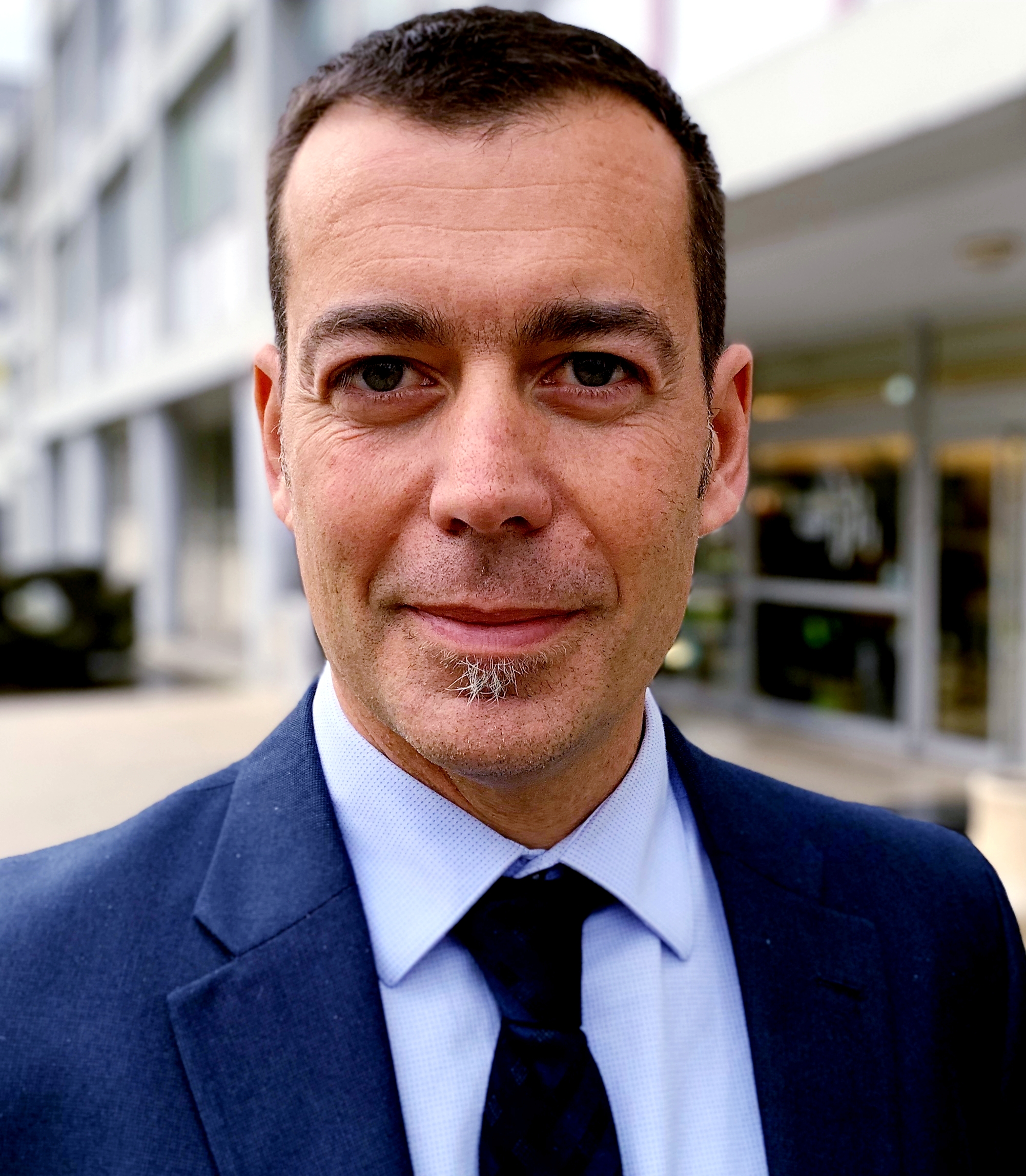}}
\noindent {\textbf{Dr. Emilio Calvanese Strinati} is the 6G Program and smart
devices and telecommunications scientific and innovation director at the French Atomic Energy Commission’s Electronics and Information Technologies Laboratory, Grenoble, France. His current research interests are in the area of beyond 5G future enabling technologies such as high frequency communications, mobile edge computing and distribute intelligence.}

\parpic{\includegraphics[width=60pt,height=70pt,clip,keepaspectratio]{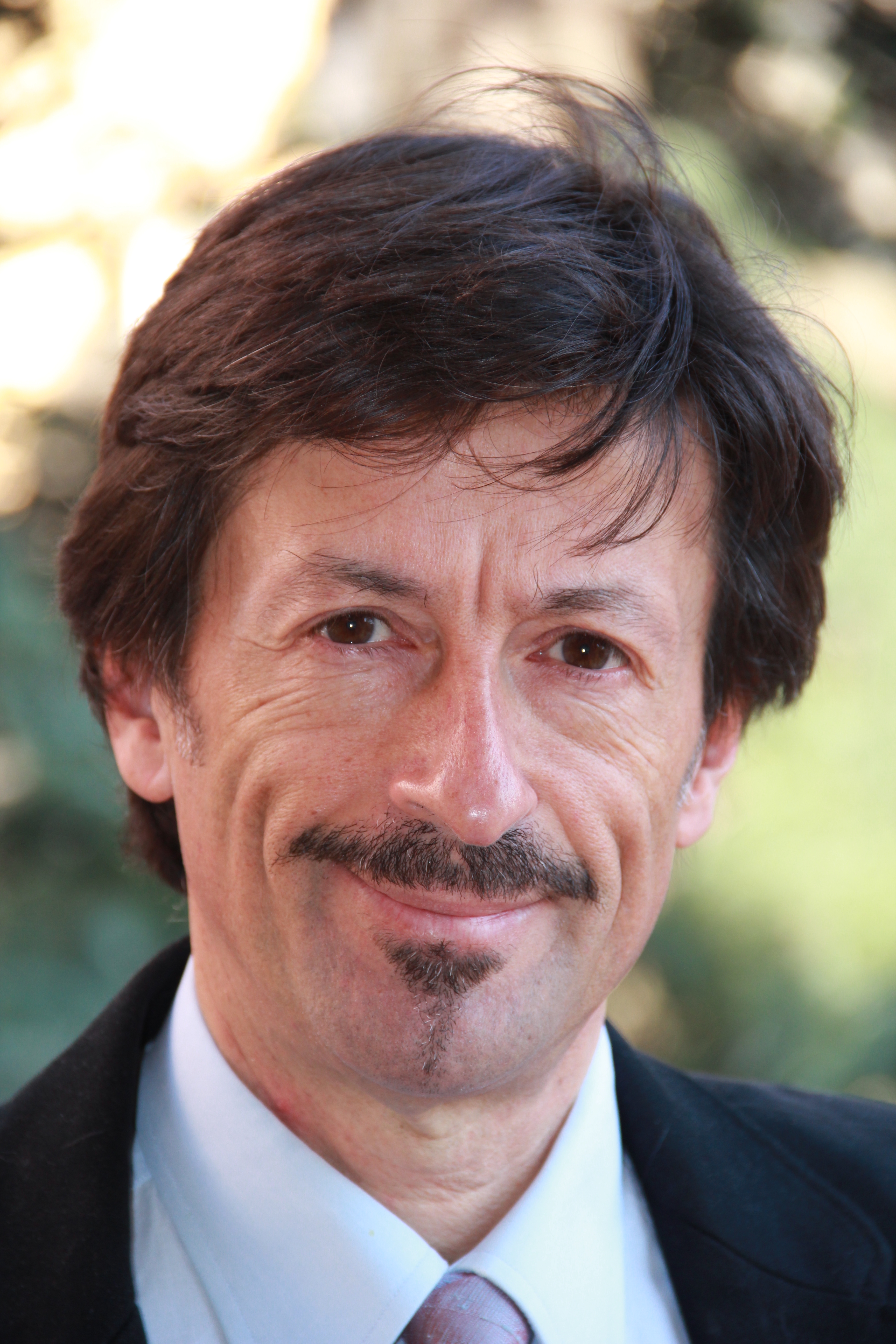}}
\noindent {\textbf{Sergio Barbarossa} is a full professor in the Department of Information Engineering,
Electronics, and Telecommunications, Sapienza University of Rome, Italy. His current research interests are in the area of millimeter-wave communications and mobile edge computing,
topological signal processing, machine learning, and distributed optimization. He is a Fellow of the IEEE and of the European Association for Signal Processing.}

\parpic{\includegraphics[width=60pt,height=70pt,clip,keepaspectratio]{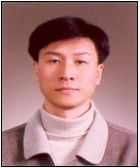}}
\noindent {\textbf{Taesang Choi} received his MS and Ph.D degrees in computer science and telecommunications in Univ. of Missouri-Kansas City in 1988 and 1995 respectively. He joined ETRI in 1996 and is currently working as a principal research staff. He has participated in the over 20 national and international projects. His research focuses on the traffic engineering, traffic analytics, management and orchestration of the SDN/NFV/5G network slice and Quantum key distribution network.}

\parpic{\includegraphics[width=60pt,height=70pt,clip,keepaspectratio]{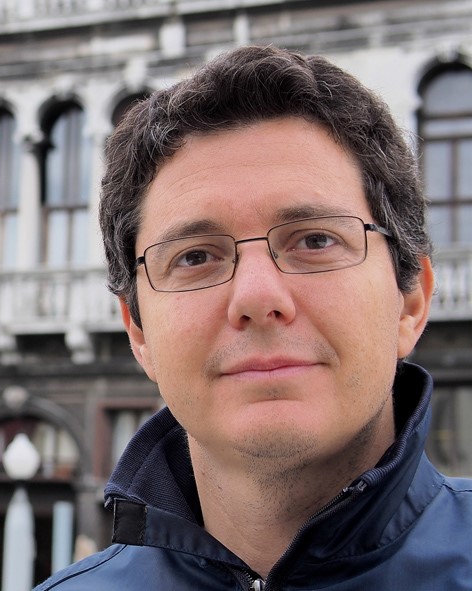}}
\noindent {\textbf{Antonio Pietrabissa} is Associate Professor at  the University of Rome “La Sapienza,” where he received his degree in Electronics Engineering and his Ph.D. degree in Systems Engineering in 2000 and 2004, respectively. Since 2000, he has participated in about 20 EU and National research projects. His research focuses on the application of systems and control theory to the analysis and control of networks. }

\parpic{\includegraphics[width=60pt,height=70pt,clip,keepaspectratio]{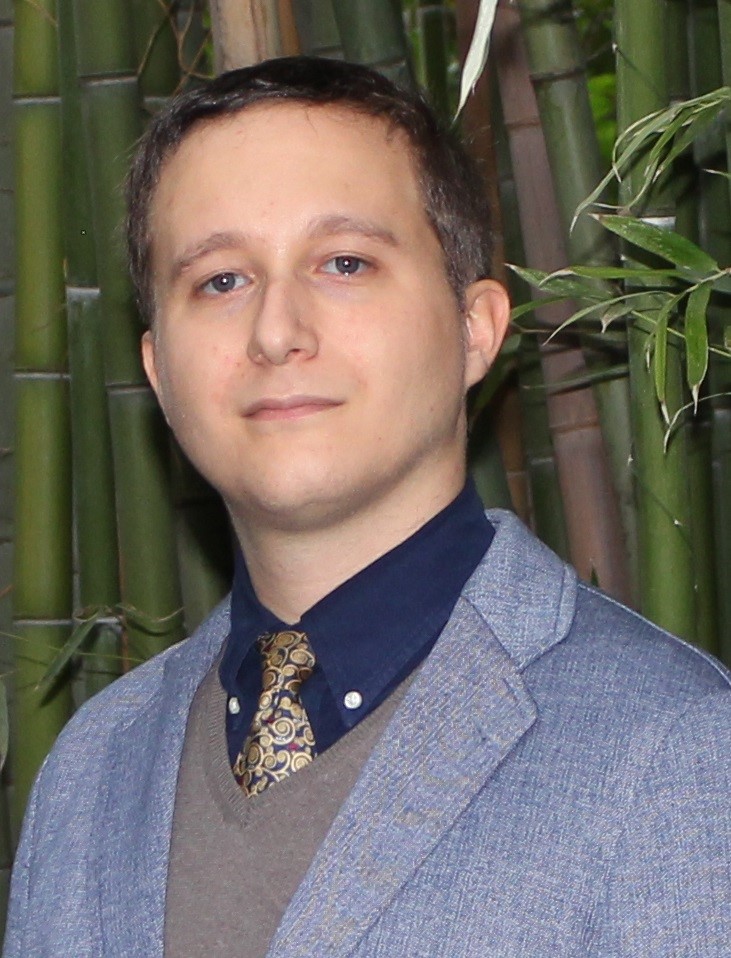}}
\noindent {\bf Alessandro Giuseppi} received from the University of Rome “La Sapienza” his M.Sc. degree in Control Engineering in 2016 and his Ph.D. in Automatica in 2019. Currently, he is a Postdoctoral Researcher in Automatic Control at the same university. Since 2016, he has participated in 5 EU and National research projects. His main research activities are in the fields of network control and intelligent systems.

\parpic{\includegraphics[width=60pt,height=70pt,clip,keepaspectratio]{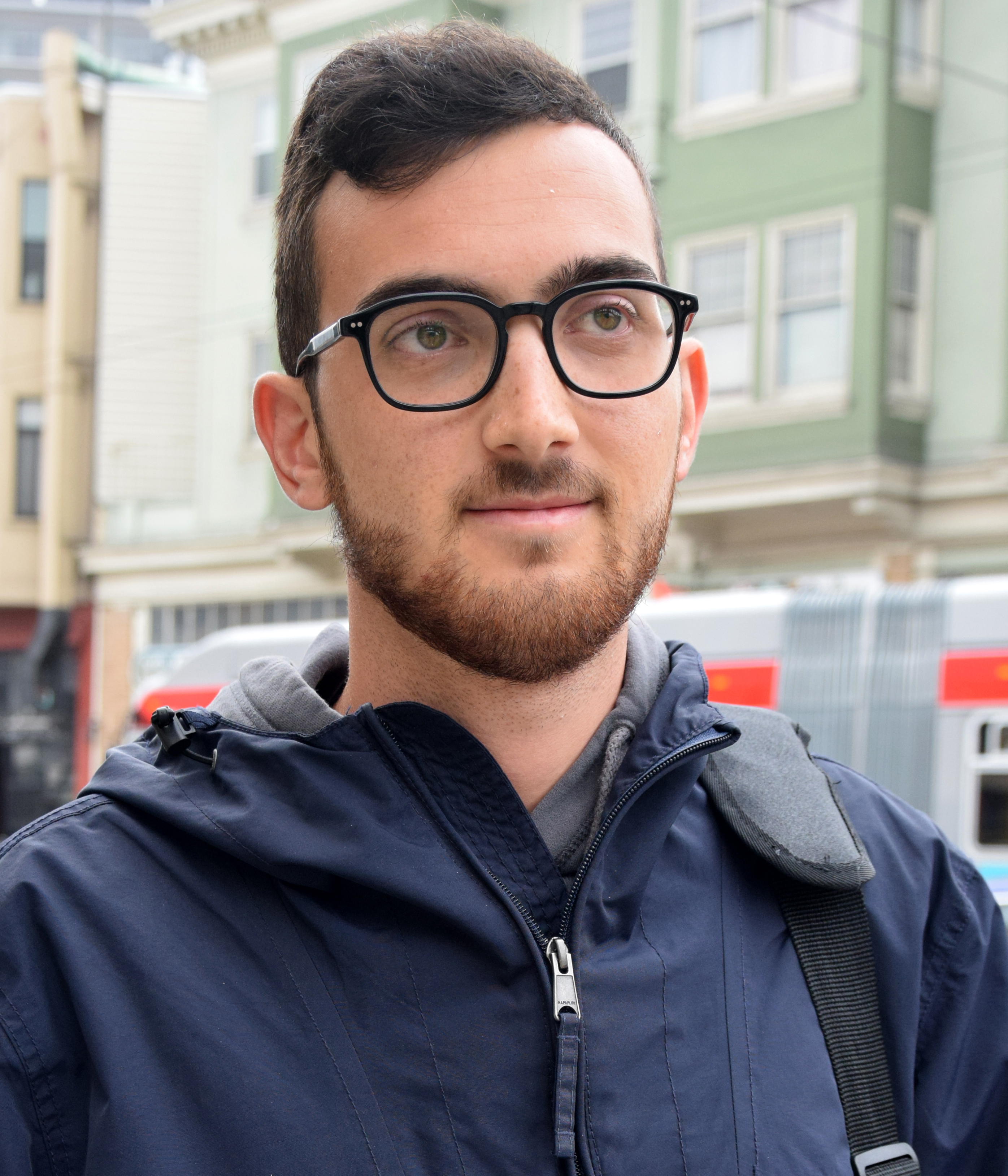}}
\noindent {\textbf{Emanuele De Santis} received the M.Sc. degree in Engineering in Computer Science in 2019 from Sapienza University of Rome, where he is currently a PhD student in Automatic Control. He is participating to H2020 projects 5G-ALLSTAR and 5G-Solutions. His main research activities are in the field of network control.}

\parpic{\includegraphics[width=60pt,height=70pt,clip,keepaspectratio]{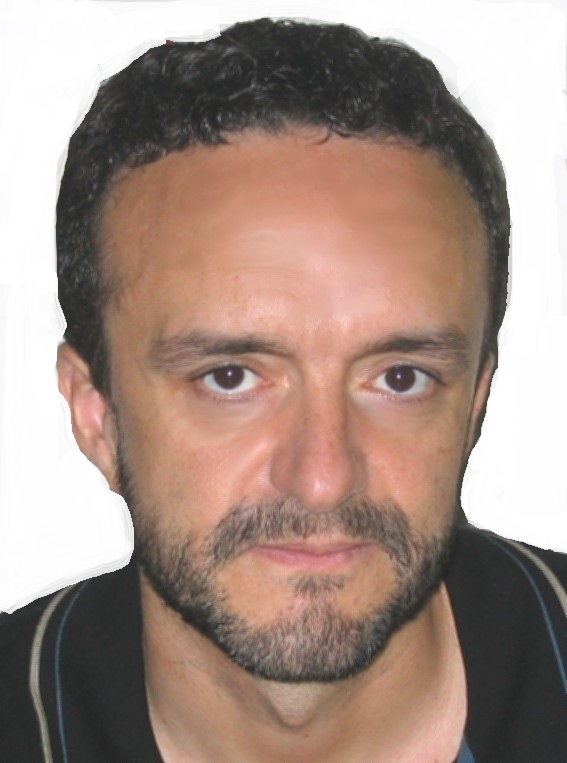}}
\noindent {\textbf{Josep Vidal} is a full professor in the Department of Signal Theory Communications, Universitat Politècnica de Catalunya, Spain. His research interests are in the areas of communications, edge computing and machine learning, with a focus in the design of solutions for beyond 5G systems. He is leading the UPC participation in the 5G Barcelona initiative.}

\parpic{\includegraphics[width=60pt,height=70pt,clip,keepaspectratio]{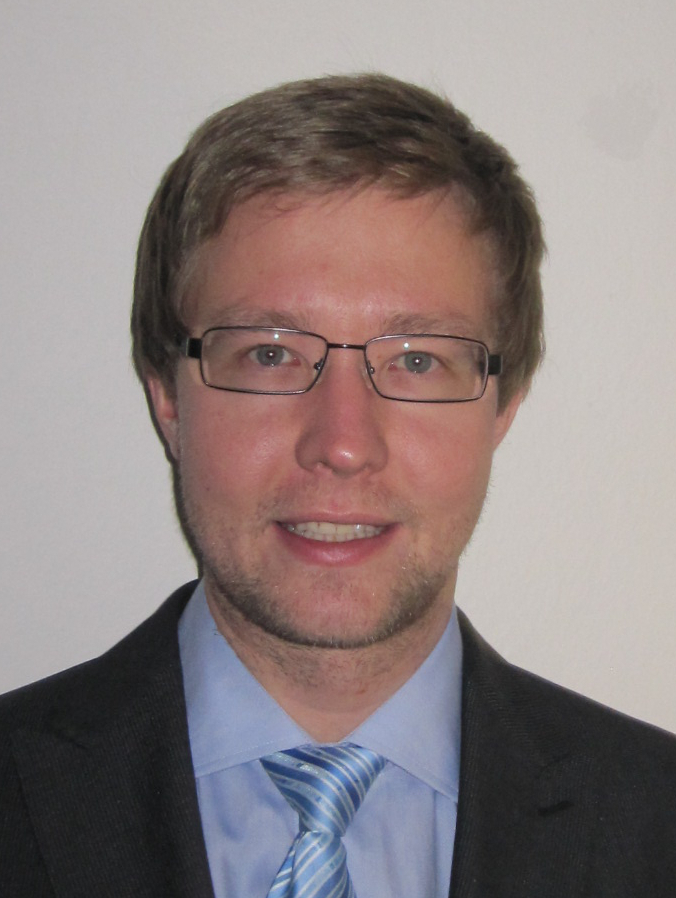}}
\noindent {\textbf{Zdenek Becvar} is Associate Professor at the Department of Telecommunication Engineering, Czech Technical University in Prague (CTU), Czech Republic. From 2006 to 2007, he joined Sitronics R\&D center. Furthermore, he was involved Vodafone R\&D center at CTU in 2009. He is focused on radio resource and mobility management in future mobile networks (beyond 5G, 6G), edge computing, and machine learning.}

\parpic{\includegraphics[width=60pt,height=70pt,clip,keepaspectratio]{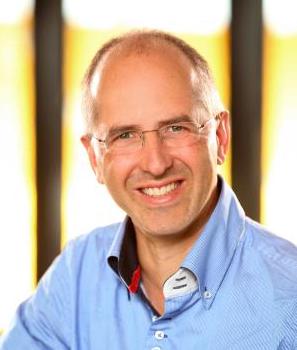}}
\noindent {\textbf{Thomas Haustein} received the  Ph.D. degree in communications from the TU Berlin, Germany, in 2006. He worked with Nokia Siemens  on 4G from 2006 to 2008. Since 2009, he is heading the Wireless Department, Fraunhofer HHI, with a focus  on 5G and industrial wireless and actively contributes to 3GPP standardization.}

\parpic{\includegraphics[width=60pt,height=70pt,clip,keepaspectratio]{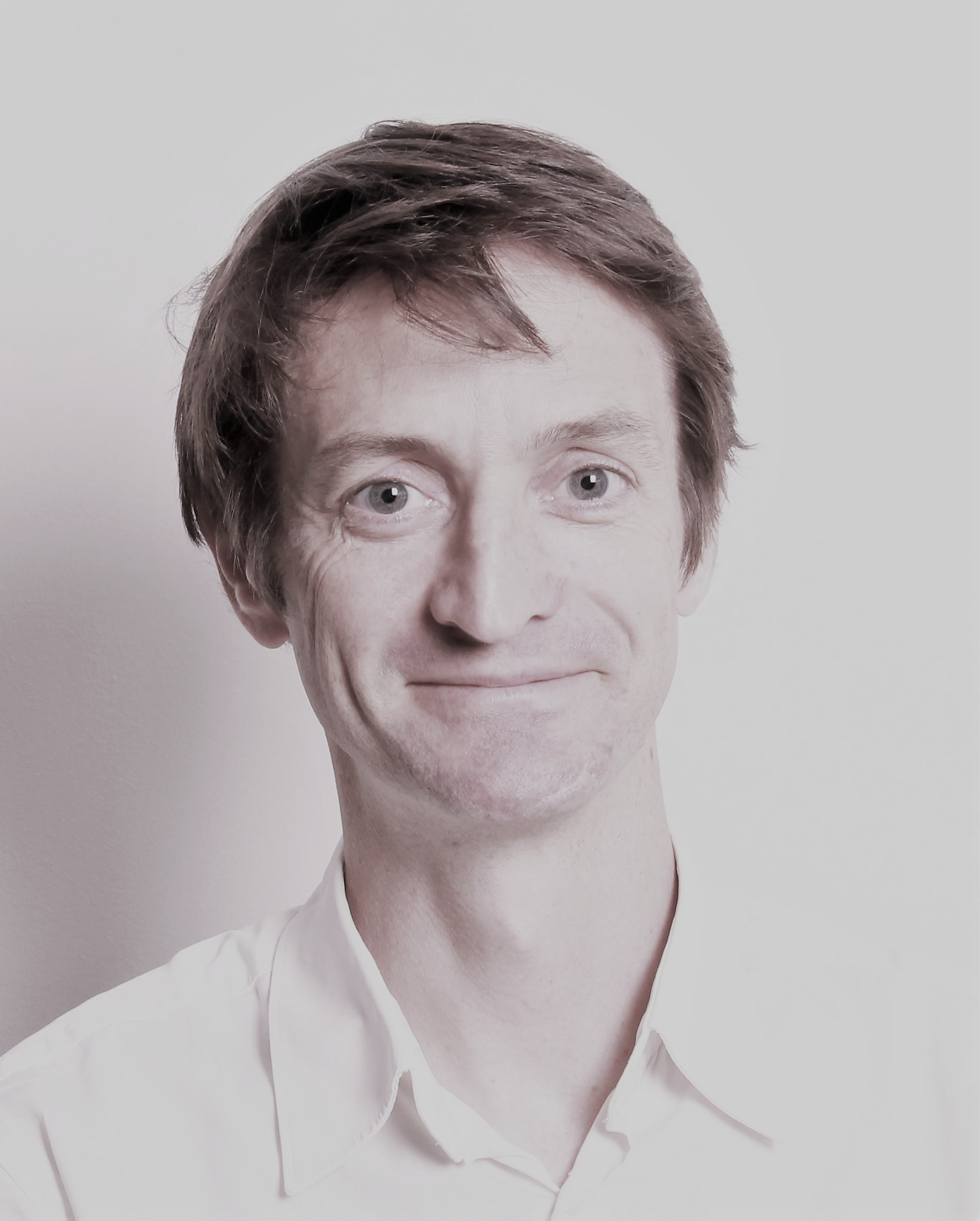}}
\noindent {\textbf{Nicolas Cassiau} is a research engineer and project manager at the French Atomic Energy Commission’s Electronics and Information Technologies Laboratory, Grenoble, France. His research interests are digital wireless communications and algorithm design, and his focus is on physical layer design and assessment for 4G/5G.}

\parpic{\includegraphics[width=60pt,height=70pt,clip,keepaspectratio]{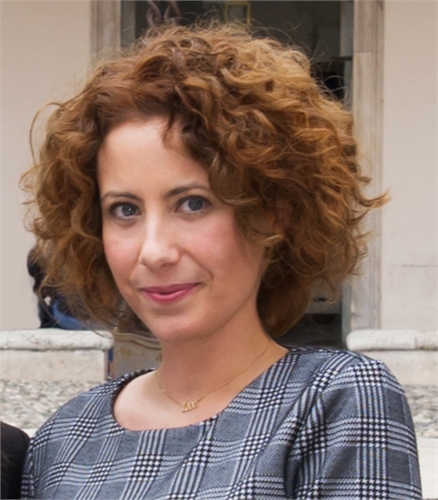}}
\noindent {\textbf{Francesca Costanzo} received the M.Sc. degree in Telecommunication Engineering in 2017 from Sapienza Univ. of Rome, where she is currently working towards the PhD degree in ICT. 
Her main research activities are in Edge Computing, 5G, and Stochastic Optimization.}

\parpic{\includegraphics[width=60pt,height=70pt,clip,keepaspectratio]{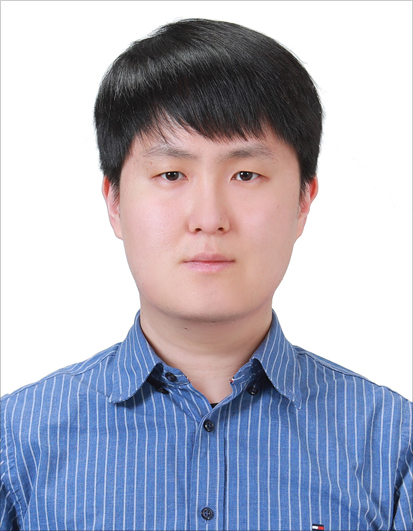}}
\noindent {\textbf{Junhyeong Kim} has been with ETRI since 2011,
where he is taking part in the development of mmWave-band vehicular communication systems.
His main research interests include millimeter-wave communications, vehicular communications, cooperative communications, and handover.}

\parpic{\includegraphics[width=60pt,height=70pt,clip,keepaspectratio]{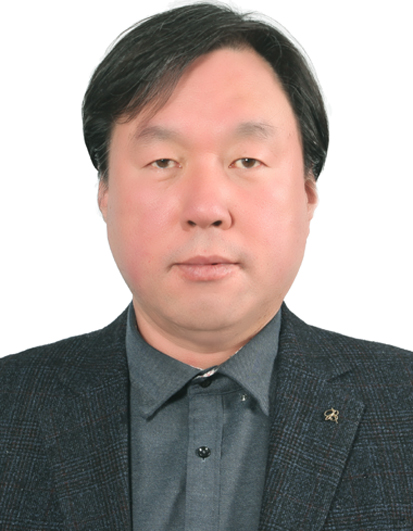}}
\noindent {\textbf{Ilgyu Kim} 
has been with ETRI since 2000,
where he involved in the standardization and development of WCDMA, HSPA and LTE.
Since 2019, he has been the assistant vice president of ETRI and managing director of Future Mobile Communication Research Division.
Currently, his main research interests include millimeter wave/THz communications and future mobile communication.}

\end{document}